\documentclass[aps,pre,onecolumn,superscriptaddress,showpacs]{revtex4}
\usepackage{amssymb}
\usepackage{epsfig}
\usepackage{graphicx}
\usepackage{float}
\usepackage{amsmath}
\usepackage{times}
\usepackage{multirow}
\usepackage{caption}
\begin{document}

\title{localized attack on clustering networks}
\author{Gaogao Dong}\email{dfocus.gao@gmail.com}
\affiliation{Nonlinear Scientific Research Center, Faculty of
Science, Jiangsu University, Zhenjiang, 212013, China}

\affiliation{Center for Polymer Studies and Department of Physics,
Boston University, Boston, MA 02215, USA}

\author{Huifang Hao}
\affiliation{Nonlinear Scientific Research Center, Faculty of
Science, Jiangsu University, Zhenjiang, 212013, China}

\author{Ruijin Du}
\affiliation{Nonlinear Scientific Research Center, Faculty of
Science, Jiangsu University, Zhenjiang, 212013, China}

\author{Shuai Shao}
\affiliation{Nonlinear Scientific Research Center, Faculty of
Science, Jiangsu University, Zhenjiang, 212013, China}

\author{H. Eugene. Stanley}
\affiliation{Center for Polymer Studies and Department of Physics,
Boston University, Boston, MA 02215, USA}

\author{Havlin Shlomo}
\affiliation{Minerva Center and Department of Physics, Bar-Ilan
University, Ramat-Gan 52900, Israel}

\begin{abstract}
Clustering network is one of which complex network attracting plenty of scholars to discuss and study the structures and cascading process. We primarily analyzed the effect of clustering coefficient to other various of the single clustering network under localized attack. These network models including double clustering network and star-like NON with clustering and random regular (RR) NON of ER networks with clustering are made up of at least two networks among which exist interdependent relation among whose degree of dependence is measured by coupling strength. We show both analytically and numerically, how the coupling strength and clustering coefficient effect the percolation threshold, size of giant component, critical coupling point where the behavior of phase transition changes from second order to first order with the increase of coupling strength between the networks. Last, we study the two types of clustering network: one type is same with double clustering network in which each subnetwork satisfies identical degree distribution and the other is that their subnetwork satisfies different degree distribution. The former type is treated both analytically and numerically while the latter is treated only numerically. In each section, we compared two results obtained from localized attack and random attack according to Shao $et$ $al.$\cite{S.Shao2013}.

\end{abstract}
%\date{\today}
\maketitle

\section{introduction}

Newman $et$ $al.$ stated the study of complex networks was in its infancy period and predicted three directions which can been researched in 2003\cite{Newman M E J2003}. Recent ten years witnesses the rapid development of complex network which is abstracted about a variety of relationships based on all kinds of field the real and defined more about the nature to precisely and comprehensively depict networks\cite{JeongH2000,Rosato V2008,Dorogotsev S N2003,Bashan A2012,Rinaldi S M2001}. A most classical example is the relationship net of movie and television stars if and only if the two stars appeared in the same work. Plenty of approaches are used to study the evolution mechanism and cascading process of network when it encounters attack or damage and obtain the theories of dynamical processes taking place on networks\cite{Newman M2010}. Attack ways which can cause recursive action and lead to a cascade failures in network by removing a fraction of nodes\cite{Buldyrev S V2010}, generally speaking, following three types are most representative: random attack\cite{Gao J2011}, localized attack\cite{S.Shao2014} and target attack\cite{G.G.Dong2013}. The varieties of complex network are also diversiform, for example, ER network, RR network and clustering network\cite{Serrano M ¨¢2006} and so on.

Here, we mainly introduce the clustering network and localized attack. Clustering network has two different structures including triangle and single lines. Especially, triangle indicates that if $A$ has a certain relation with $B$ and $C$, there will have a high probability to exist this kind of relationship between $B$ and $C$, for example, friendship, business relation and cooperative relationship and so on\cite{Newman M E2001}. The probability is defined as the clustering coefficient of clustering network. Newman $et$ $al.$ proposed a random-graph model of a clustered network and obtained component sizes, existence and size of a giant component by extending the generating function way, a widely used theoretical tool in complex network \cite{M.E.J.Newman2009}. Moreover, Gao $et$ $al.$ studies the model of Network of Networks and other models\cite{Gao J2011}. Later, Dong $et$ $al.$ made a deep research on percolation behaviors of partial support-dependent networks\cite{Dong G2014}. Shao $et$ $al.$ studied different structure of complex networks under localized attack which means the nodes were removed from a certain area\cite{S.Shao2014}. However, the detailed cascading process about incorporating clustering networks with localized attack has not been discussed and explored.

According to contents, point of view and results of the aforementioned paper, this paper was broken down into four sections: (i)Single clustering network. (ii) Two interdependent networks with clustering. (iii) Network of networks with clustering. (iv) Fixed degree distribution. For all conditions, we also show the results of random attack\cite{S.Shao2013}. In addition, we study the relationship among average degree $\langle k \rangle$, clustering coefficient $c$, critical coupling $q_c$ and the size of giant component $P_{\infty}$ and the robustness of network. We primarily deduce a set of formulas and analyze the size of giant component and the effect of clustering coefficient to critical threshold and average degree to coefficient. Then, we generalize the results to the latter sections and adjust the solution according to the network structure. Section (iii) includes two model, star-like NON with clustering and random regular(RR) NON of ER networks with clustering. Finally, section (iv) shows the difference between fixed degree distribution (FDD) and doubly Poisson distribution (DPD) which are two classes of degree distribution of clustering network.

\section{single clustering network}

For any node with degree $k$ in a single clustering network, it is the sum of two types of links number, the number of single links $s$ and the number of triangle $t$, i.e., $k=s+2t$.
A single clustering network satisfies double Poisson distribution
\begin{equation}\label{eq1}
p_{s,t}=e^{\langle s \rangle}\frac{\langle s \rangle^s}{s!}e^{\langle t \rangle}\frac{\langle t \rangle^t}{t!}.
\end{equation}
where $\langle s \rangle$ means average number of single lines, $\langle t \rangle$ means average number of triangles and  average degree $\langle k \rangle$ meet $\langle k \rangle=\langle s \rangle+2\langle t \rangle$. The more triangles, the more clustering of networks. So the clustering coefficient is defined as $c=\frac{2 \langle t\rangle}{2 \langle t \rangle + \langle k \rangle^2}$\cite{Newman M E J2001}. The corresponding generating function is defined as\cite{M.E.J.Newman2009}
\begin{equation}\label{eq2}
G_0(x,y)=\sum_{s,t=0}^\infty p_{st}x^sy^t=e^{\langle s \rangle(x-1)}e^{\langle t \rangle(y-1)},
\end{equation}
First, randomly choose a node as "root" node. Second, we range all nodes of network shell by shell which defined as the set of nodes according to the distance, i.e., shell $l$ in network includes those nodes whose distance to the given root node are entirely equal $l$ and who are positioned randomly and coequally\cite{Shao J2009}. Then beginning to attack this clustering network use the same way of localized attack as expression in Shao $et$ $al.$ where the process of localized attack is separated into following two steps\cite{S.Shao2014}. We start with removing a $1-p$ fraction of nodes from the root node shell by shell. In this process , notes that we only remove the links connected by both removed nodes but keep the links between the removed nodes and the remaining nodes. The generating function of the remaining network is
\begin{equation}\label{eq3}
G_1(x,y)=\frac{G'_0(hx,h^2y)}{G'_0(h,h^2)}=\frac{e^{\langle s \rangle(hx-1)}e^{\langle t \rangle(h^2y-1)}}{e^{\langle s \rangle(h-1)}e^{\langle t \rangle(h^2-1)}},
\end{equation}
where $G_0(h,h^2)=p$. Here and now, the proportion of a link end at
an unremoved node for the remaining network can be expressed as
\begin{equation}\label{eq4}
\tilde{p}=\frac{G'_0(f,f^2)}{G'_0(1,1)f}=\frac{p}{f},
\end{equation}
Second step considers removing this kind of links which are connecting the removed nodes to the remaining nodes. At the end of this attack, the generating function of the remaining network is equal to
\begin{equation}\label{eq5}
\begin{split}
& G^p_0(x,y)=G_1(1-\tilde{p}+\tilde{p}x,1-\tilde{p}+\tilde{p}y),
\end{split}
\end{equation}
By combining $\tilde{G}_0(1-p+px,1-p+py)=G^p_0(x,y)$ and Eq.(5), we deduce
\begin{equation}\label{eq6}
\tilde{G}_0(x,y)=G_1(1+\frac{\tilde{p}}{p}x-\frac{\tilde{p}}{p},1+\frac{\tilde{p}}{p}y-\frac{\tilde{p}}{p}),
\end{equation}
After plugging Eq.(6) into Eq.(3) and simplifying, we can get
\begin{equation}\label{eq7}
\tilde{G}_0(x,y)=e^{\langle s \rangle(x-1)}e^{\langle t \rangle h(y-1)},
\end{equation}
The fraction of nodes that belong to the giant component of the remaining single clustering network is
\begin{equation}\label{eq8}
g(p)=1-\tilde{G}_0[1-p(1-f(p)),(1-p(1-f(p)))^2],
\end{equation}
where $f(p)$ satisfies $f(p)=\tilde{G}_1[1-p(1-f(p),(1-p(1-f(p))^2)]$ and $\tilde{G}_1(x,y)=\frac{\tilde{G}'_0(x,y)}{\tilde{G}'_0(1,1)}=\tilde{G}_0(x,y)$.
The final fraction of giant component for the original clustering network is $P_{\infty}$ (as shown in Fig.1(a)), which satisfies
\begin{equation}\label{eq9}
P_{\infty}=pg(p).
\end{equation}
By adding condition $f\rightarrow0$ to equation
\begin{equation}\label{eq10}
f(p)=e^{\langle s \rangle(1-p(1-f(p)-1)}e^{\langle t \rangle h((1-p(1-f(p)))^2-1)}.
\end{equation}
The critical threshold of second order phase transition $p_c$ is solved as
\begin{equation}\label{eq11}
p_c=\frac{1}{\langle s \rangle+2 h \langle t \rangle},
\end{equation}
where $h=G_0^{-1}(p_c)$. In addition , there exist relation between $c,k$ and $\langle s \rangle,\langle t \rangle$, so we draw up the Fig.2 (a) and (b). By comparing results from localized attack and randomly attack, it is obvious that the bigger clustering coefficient $c$ corresponds to a bigger phase point, but the smaller degree $k$ corresponds to a bigger phase point. On the other hand, the strength of localized attack is more strong than random attack.
\begin{figure}[H] \setlength{\abovecaptionskip}{-0.01cm} \centering
\scalebox{0.50}[0.50]{\includegraphics{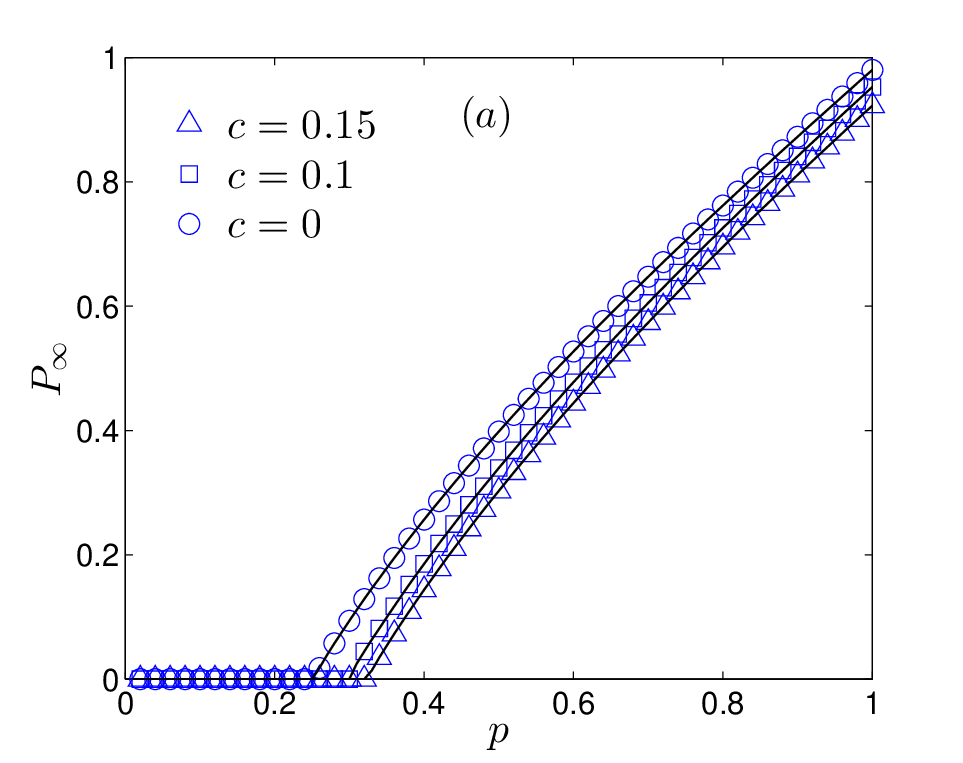}}\centering
\caption{(a) Size of giant component $P_{\infty}$, as a function of $p$ with parameter $\langle k \rangle=4$ for different clustering coefficient, where symbols are from numerical solution and solid lines are from simulations with network nodes $N=10000$. The two kinds of results are agree well with each other. The phase transition point increases with the increase of $c$. Note that the behavior of phase transition is continuous, i.e, second order.}
\end{figure}

\begin{figure}[H] \setlength{\abovecaptionskip}{-0.01cm} \centering
\scalebox{0.5}[0.52]{\includegraphics{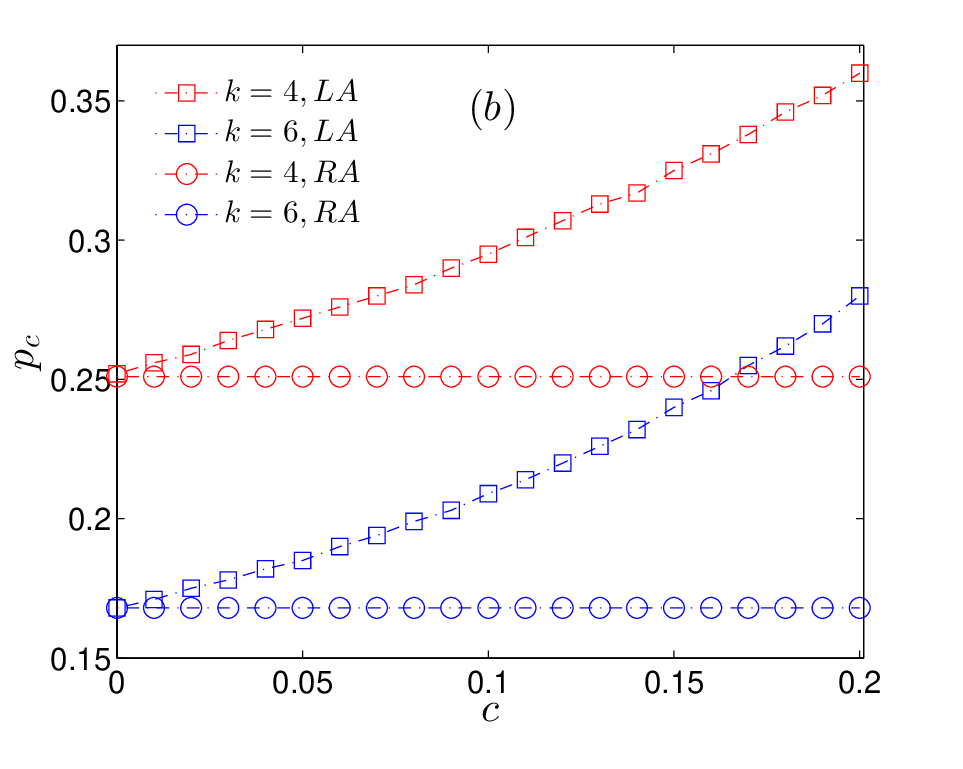}}
\scalebox{0.5}[0.5]{\includegraphics{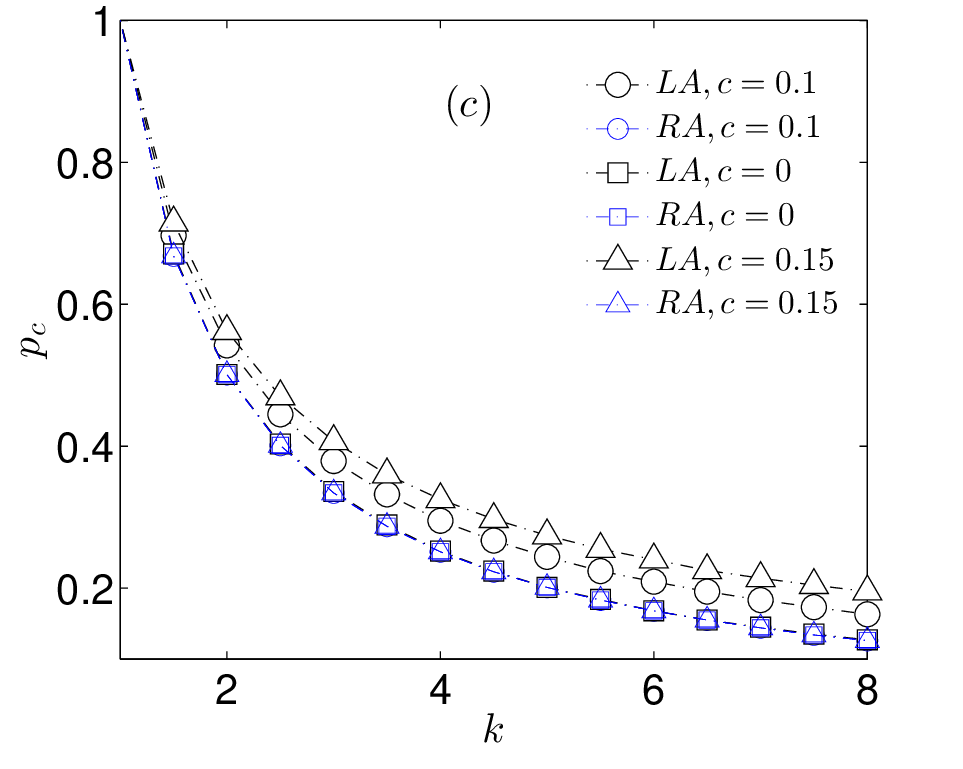}}\centering
\caption{(b) Percolation threshold $p_c$ as a function of clustering coefficient $c$ with different $\langle k \rangle$ ($\langle k \rangle=4,6$). The square represent result of local attack (LA) and circle represent random attack (RA). (c) Percolation threshold  $p_c$ as a function of $k$ with different clustering coefficient $c$ ($c=0,0.1$ $and$ $0.15$). The black denotes the results of RA and the black denote LA.}
\end{figure}

\section{double clustering network}
This model includes two clustering networks $A$ and $B$ which respectively obey a double Poisson distribution as the same with single clustering network,that is to say, $\langle k \rangle_A=\langle k \rangle_B=\langle k \rangle$, $\langle s \rangle_A=\langle s \rangle_B=\langle s \rangle$, $\langle t \rangle_A=\langle t \rangle_B=\langle t \rangle$. Assume a $q_A$ fraction of nodes in network $A$ depend on nodes in network $B$ and a $q_B$ fraction of nodes of network $B$ also depend on nodes in network $A$. It means if a nodes in network $B$ which depends on the other failure node in network $A$, this node will fail, and vice versa. Notes that their relation of two networks' nodes is one to one and no-feedback condition. We start with removing a $p$ fraction of nodes from network $A$ and network $B$ separately by localized attack and the system finally get steady state after several iterations. At this time, the remaining fraction of nodes in network $A$ and network $B$ are equal to $X$ and $Y$\cite{G.G.Dong2013},
\begin{equation}\label{eq4}
\begin{split}
& X=p(1-q_A(1-g_B(Y))p),\\
& Y=p(1-q_B(1-g_A(X))p),\\
\end{split}
\end{equation}
The size of giant components of two network $A$ and $B$ can be expressed as $P_{\infty,1}$ and $P_{\infty,2}$ (as shown in Fig.3 (a))
\begin{equation}\label{eq2}
P_{\infty,1}=Xg_A(X),P_{\infty,2}=Yg_B(Y).
\end{equation}
where $g_A(X)$ and $g_B(Y)$ satisfy
\begin{equation}\label{eq4}
\begin{split}
& g_A(X)=1-e^{\langle s \rangle(1-X(1-f_A(X))-1)}e^{\langle t \rangle h((1-X(1-f_A(X)))^2-1)},\\
& g_B(Y)=1-e^{\langle s \rangle(1-Y(1-f_B(Y))-1)}e^{\langle t \rangle h((1-Y(1-f_B(Y)))^2-1)}.\\
\end{split}
\end{equation}
and
\begin{equation}\label{eq4}
\begin{split}
& f_A(X)=e^{\langle s \rangle(1-X(1-f_A(X))-1)}e^{\langle t \rangle h((1-X(1-f_A(X)))^2-1)},\\
& f_B(Y)=e^{\langle s \rangle(1-Y(1-f_B(Y))-1)}e^{\langle t \rangle h((1-Y(1-f_B(Y)))^2-1)}.\\
\end{split}
\end{equation}
where $h=G_0^{-1}(p)$.

Because of the intricate iterative formulas, we have not obtained explicit expressions of the critical threshold of phase transition and only provide numerical solution as shown in Fig.3 (b),(c) and (d).
Fig.3 (a) shows size of the giant component in network $A$ after the cascading process finished . The phase transition point changes from second order ($q=0.2$) to first order ($q=0.8$). The network becomes more and more robust with the increase of clustering coefficient $c$ from $c=0$ to $c=0.15$. In each curve, simulation results show great agreement with the theoretical results obtained from Eqs.(13). By combining Eqs.(10),(12),(13) and (14), we can obtain the image solution about $p_c$ as a function of $q$, as Fig.3 (b) shows, increasing coupling strength will reduce the robustness of network. There are a critical coupling $q_c$ to distinguish the class of phase transition for a certain set of $c$ and $k$, and this value almost remains the same for different $c$ which means the effect of clustering coefficient to coupling strength is nearly neglected for localized attack and random attack. Moreover, average degree $\langle k \rangle$ and clustering coefficient $c$ also influence the critical threshold (see Fig.4 (a) and (b)).
For a fixed coupling strength, bigger clustering coefficient and degree make networks less robust. While for weak coupling, the difference between of localized attack and random attack on robustness is bigger.

\begin{figure}[H] \setlength{\abovecaptionskip}{-0.05cm} \centering
\scalebox{0.5}[0.5]{\includegraphics{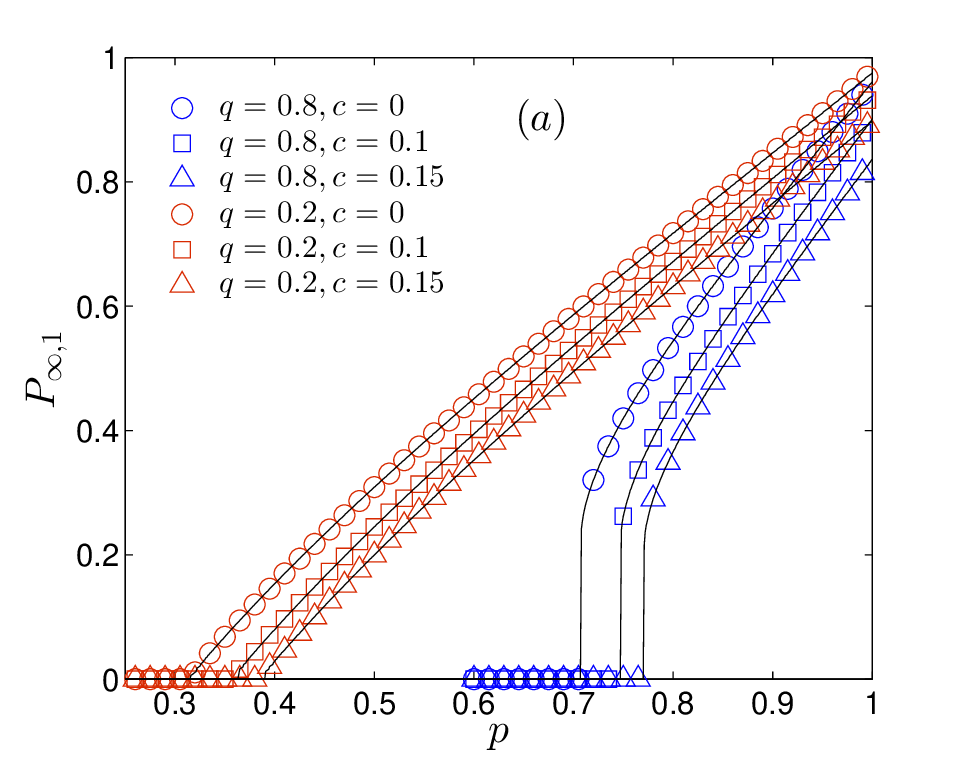}}
\scalebox{0.5}[0.5]{\includegraphics{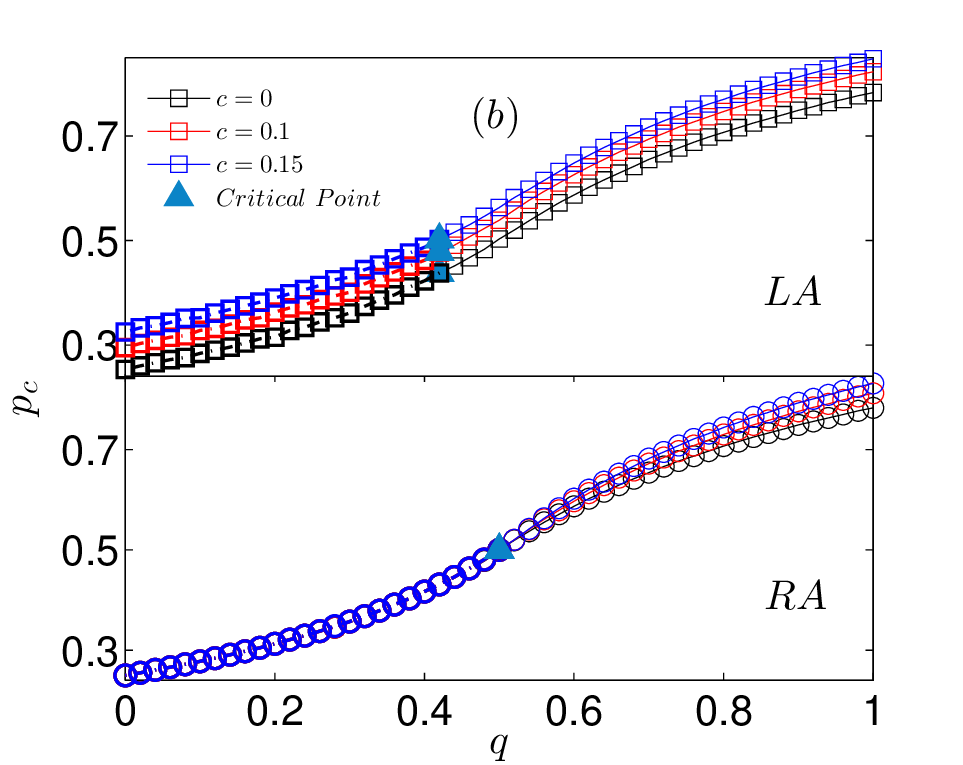}}\centering

\caption{ (a) Size of giant components $P_{\infty,1}$ as a function of $p$ for $\langle k \rangle=\langle k \rangle_A=\langle k \rangle_B=4$, where solid lines indicate the solution of theoretical predictions and symbols indicate simulations with network size $N=10000$ under different coupling strength $q$ and clustering coefficient $c$ ($0, 0.1$ $and$ $0.15$). For strong coupling ($q=0.8$), the sizes of giant component of each network changes abruptly at critical threshold $p_c$ where shows a first order phase transition behavior. For weak coupling ($q=0.2$), the behavior is continuous, which means second order phase transition. (b) Critical threshold $p_c$ as a function of interdependent strength $q$ ($q=q_A=q_B$) for $c=0, 0.1$ $and$ $0.15$ with $\langle k \rangle=4$. For a certain $\langle k \rangle$ and $c$, there exists a critical point $q_c$ (full triangle) Dashed lines denote second order phase transition (below the full blue triangle) and solid lines denote first order phase transition (above the full blue triangle).}
\end{figure}
\begin{figure}[H] \setlength{\abovecaptionskip}{-0.05cm} \centering
\scalebox{0.5}[0.5]{\includegraphics{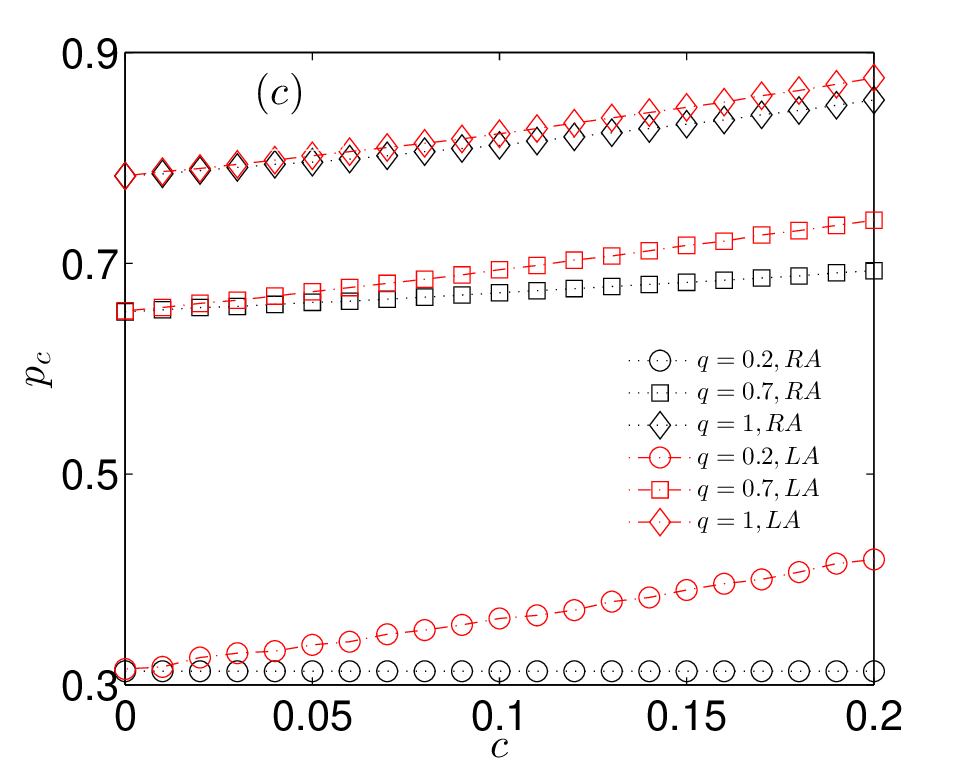}}
\scalebox{0.5}[0.5]{\includegraphics{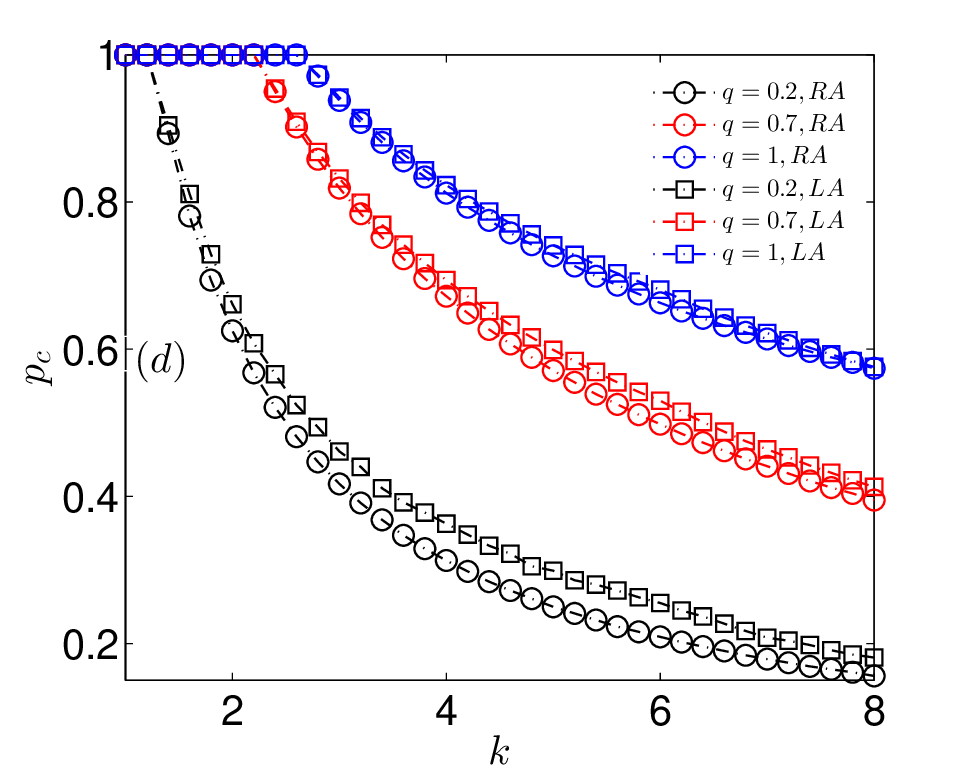}}\centering

\caption{ (c) Critical threshold $p_c$ as a function of $c$ for $q=0.2,0.7$ $and$ $1$ with $\langle k \rangle=\langle k \rangle_A=\langle k \rangle_B=4$. The black symbol lines denote the solution from RA and the red denote it from LA. (d) Critical threshold $p_c$ as a function of average degree $\langle k \rangle$ for $q=0.2,0.7$ $and$ $1$ with $c=0.1$. Circle lines denote the solution from RA and square lines denote it from LA. For (c) and (d), values of strong coupling ($q=1$) is both bigger than weak coupling ($q=0.2$) under random attack and localized attack.}
\end{figure}

\section{Network of network with clustering}
We generalize the result of double interdependent networks with
clustering to the system whose external structure is a network and
each internal node in the network is substituted into clustering
network, i.e., Network of Network(NON). For the sake of simple, we
book all the internal clustering network meet the same degree
distribution with average degree $\langle k \rangle$. Furthermore,
these internal networks partially depend on each other based on some
sort of regulation and there are not feedback in the cascading
process after suffering from attack. Here we mainly introduce two
kinds of models: A. Star-like NON with clustering B. Random
regular(RR) NON of ER networks with clustering.

\subsection{Star-like NON with clustering}
Star-like NON consists of $n$ clustering networks in which has a "center" network connected with other $n-1$ networks that are mutually independent and disconnected with each other. We assume that $q_{i,0}$ denotes the fraction of nodes in network $i$ depending on the fraction of nodes in center network $0$ ($i=1,2,3,...,n-1$). One of interdependent nodes become invalid, other nodes depending on it also lose function. The initial attack is exerted for every network by removing a fraction $1-p$ of nodes and this damage spreads in this system back and forth until this process finished and the survived network maintain stability. For simplicity , we set $q_{1,0}=q_{2,0}=...=q_{n-1,0}=q$. The remaining nodes of center network is equal to $X$ and the remaining nodes of other $n-1$ network is equal to $Y$, as following expressions [4]:
\begin{equation}\label{eq8}
\begin{split}
& X=p(1-q+pqg_B(Y))^{n-1},\\
& Y=p(1-q+pqg_A(X)[1-q+pqg_B(Y)]^{n-2}).\\
\end{split}
\end{equation}
The size of giant components of two class of network $A$ (center
network) and $B$ (other $(n-1)$ network) can be expressed as
$P_{\infty,1}$ and $P_{\infty,2}$.
\begin{equation}\label{eq2}
P_{\infty,1}=Xg_1(X),P_{\infty,2}=Yg_2(Y).
\end{equation}
where $g_1(X)$ and $g_2(Y)$ satisfy
\begin{equation}\label{eq4}
\begin{split}
& g_1(X)=1-e^{\langle s \rangle(1-X(1-f_1(X))-1)}e^{\langle t \rangle h((1-X(1-f_1(X)))^2-1)},\\
& g_2(Y)=1-e^{\langle s \rangle(1-Y(1-f_2(Y))-1)}e^{\langle t \rangle h((1-Y(1-f_2(Y)))^2-1)}.\\
\end{split}
\end{equation}
and
\begin{equation}\label{eq4}
\begin{split}
& f_1(X)=e^{\langle s \rangle(1-X(1-f_1(X))-1)}e^{\langle t \rangle h((1-X(1-f_1(X)))^2-1)},\\
& f_2(Y)=e^{\langle s \rangle(1-Y(1-f_2(Y))-1)}e^{\langle t \rangle h((1-Y(1-f_2(Y)))^2-1)}.\\
\end{split}
\end{equation}
where $h=G_0^{-1}(p)$.

Fig.6 (a) shows the size of the giant component of center network for $n=5$ comparing two sets of data $q=0.55,0.15$ and $c=0,0.1,0.15$. It is obvious that the simulation results agree will with theoretical results. For fixed $q$, the star-like NON with larger clustering coefficient is less robust, but the gap between critical points of different $c$ is very small. For fixed $c$, the phase transition changes from second order to first order with the increase of coupling strength, i.e., strong coupling strength can reduces the robustness of network. Generally speaking, the effect of coupling strength to robustness of network is more powerful than the effect of clustering coefficient. Fig.6 (b) shows the influence of coupling strength to phase transition point under localized attack and random attack. The network both become less robust with the more and more strong coupling strength. However, the critical coupling $q_c$ where the behavior of phase transition changes from first order to second order varies slightly even nearly keep unchanged. Fig.7 (c) and (d) shows the gap of critical threshold $p_c$ between localized attack and random attack become more and more clear with the increase of average degree $\langle k \rangle$ and clustering coefficient $c$, but their effect is same to the robustness of two different attack ways. For a fix $q$, the localized attack strength is stronger than random attack. Moreover, the robustness of network will be reduced with the increase of coupling strength and the decrease of average degree.  Fig.8 (e) shows the critical coupling $q_c$ in the center network for $\langle k \rangle=4,5$. With the number of subnetwork $n$ increasing, $q_c$ decreases with the increase of average degree of the network.
\begin{figure}[H] \setlength{\abovecaptionskip}{-0.05cm} \centering \scalebox{0.5}[0.5]{\includegraphics{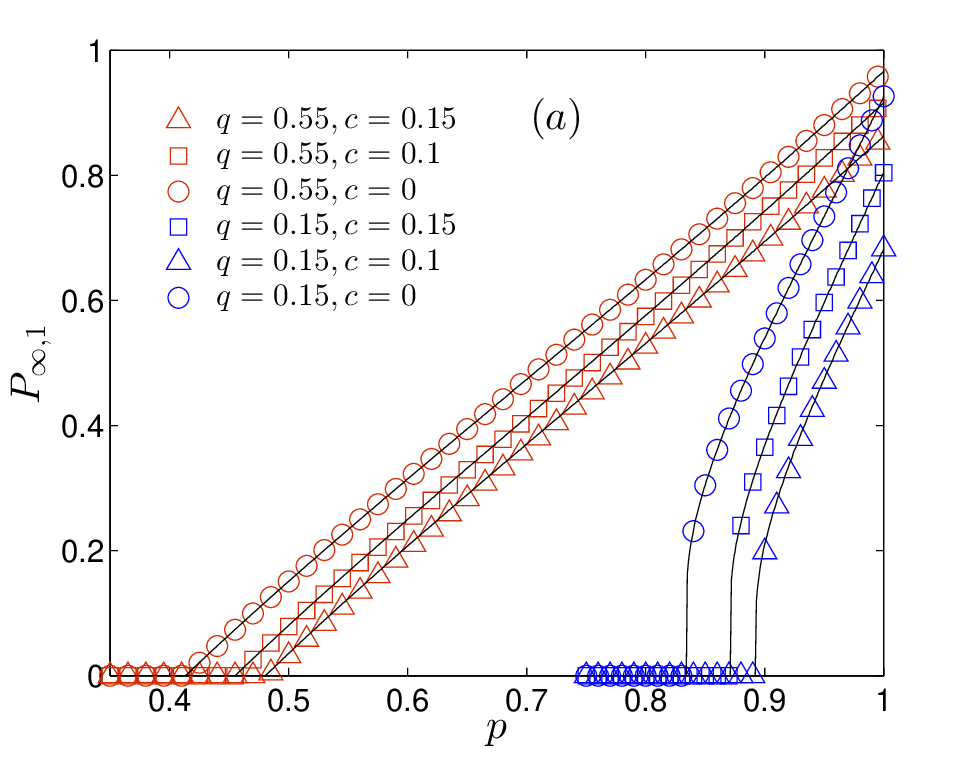}} \scalebox{0.5}[0.5]{\includegraphics{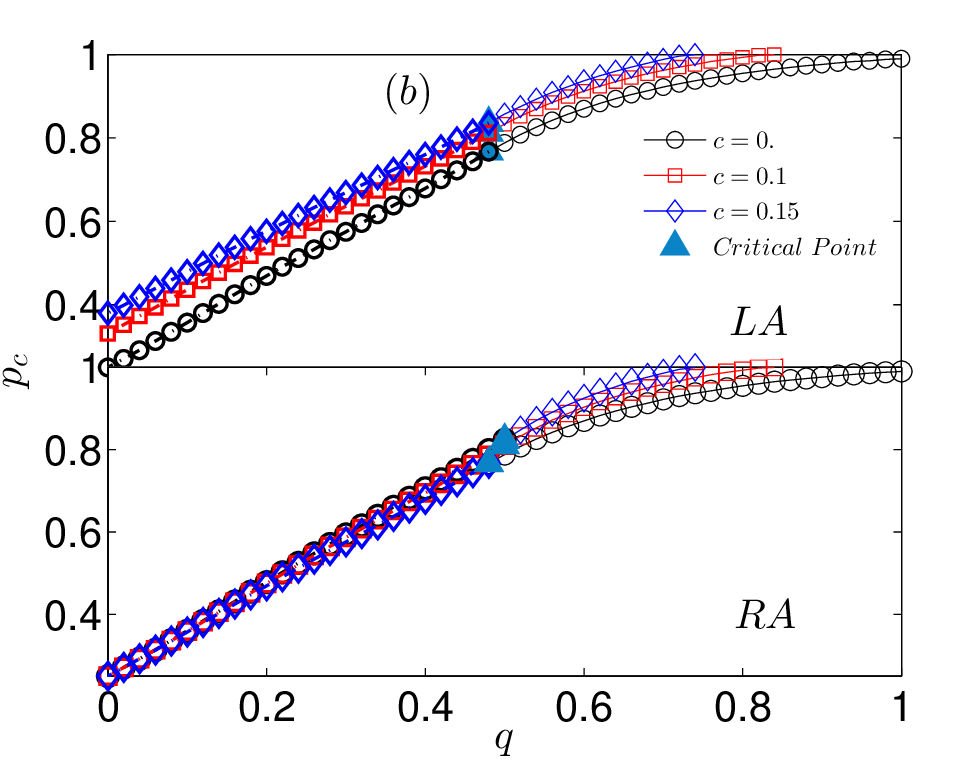}}\centering
\caption{ (a) Size of giant components $P_{\infty}$ as a function of $p$ for $c=0,0.1,0.15$ and $q=0.15,0.15$ with $\langle k \rangle=\langle k \rangle_A=\langle k \rangle_B=4$ and $n=5$, where solid lines indicate the solution of theoretical predictions and symbols indicate simulations with network size $N=10000$. For strong coupling strength ($q=0.55$), the sizes of giant component of each network changes abruptly at critical threshold $p_c$ where shows a first order phase transition behavior. For weak coupling strength ($q=0.15$), the behavior is continuous, which means second order phase transition. (b) Percolation threshold $p_c$ as a function of coupling strength $q$ for star-like NON for $c=0, 0.1$ $and$ $0.15$ with $\langle k \rangle=\langle k \rangle_A=\langle k \rangle_B=4$ and $n=5$. For each $c$, there exist a critical coupling strength $q_c$ as the full blue triangle shown indicating the behavior of phase transition is going to change from second order to first order. Dashed lines denote second order phase transition (below the full blue triangle) and solid lines denote first order phase transition (above the full blue triangle).}
\end{figure}

\begin{figure}[H]
\setlength{\abovecaptionskip}{-0.05cm} \centering
\scalebox{0.5}[0.5]{\includegraphics{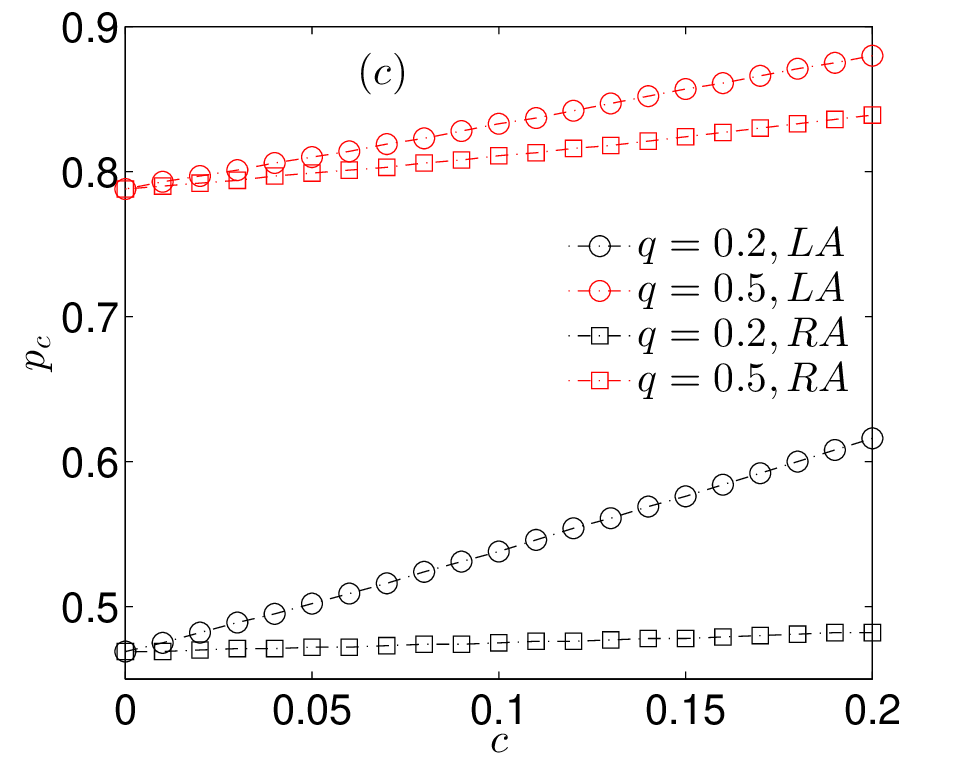}}
\scalebox{0.5}[0.5]{\includegraphics{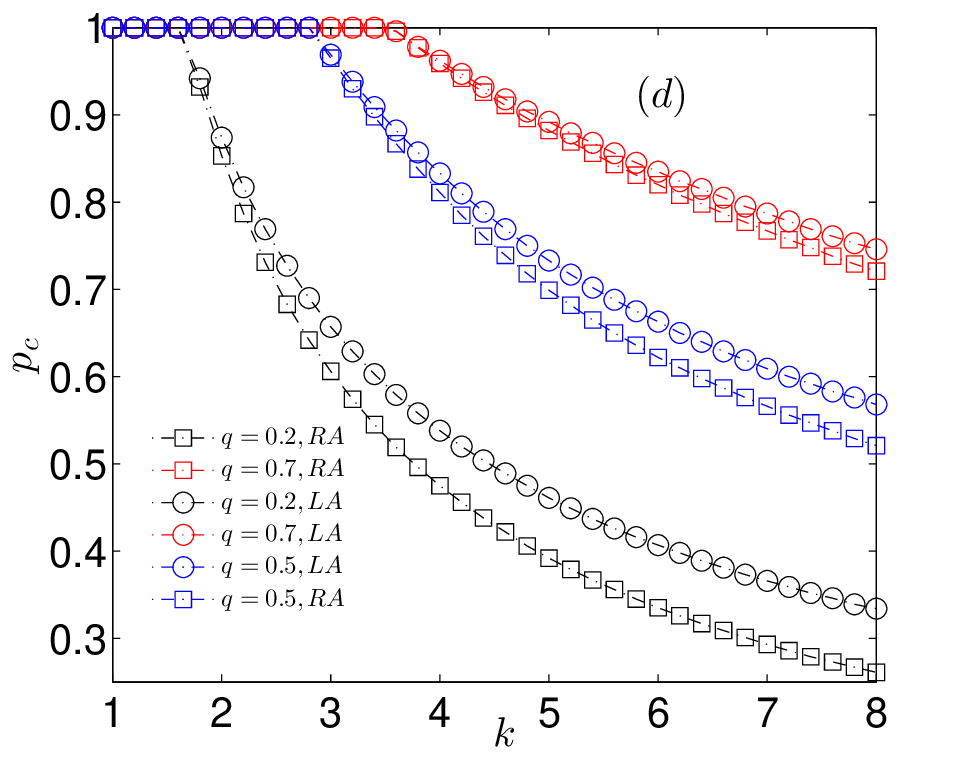}}\centering
\caption{ (c) Percolation threshold $p_c$ as a function of $c$ for $q=0.2$ $and$ $0.7$ with $\langle k \rangle=4$ and $n=5$. Circle lines denote the solution from LA and square lines denote it from RA. (d) Percolation threshold $p_c$ as a function of average degree $k$ for $q=0.2$ $and$ $0.7$ with $c=0.1$ and $n=5$. Circle lines denote the solution from LA and square line denote it from RA. For (c) and (d), values of strong coupling ($q=0.7$) is both bigger than weak coupling ($q=0.2$) under random attack and localized attack.}
\end{figure}

\begin{figure}[H]
\setlength{\abovecaptionskip}{-0.05cm} \centering
\scalebox{0.5}[0.5]{\includegraphics{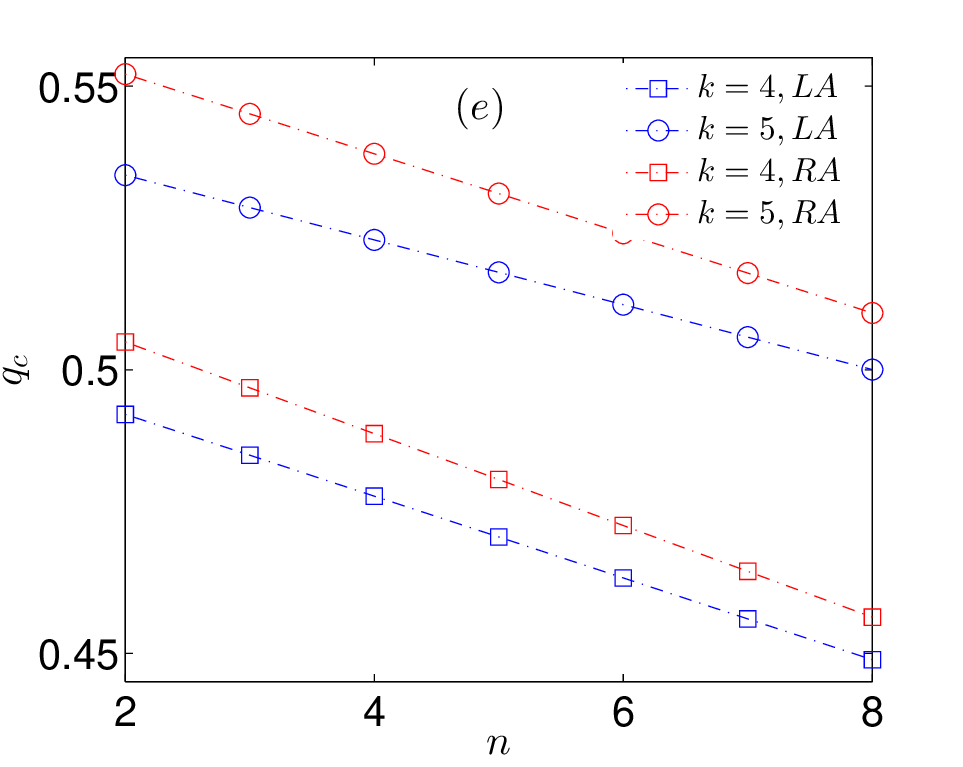}}
\centering
\caption{ (e) Critical point where phase changes from second order phase transition to first order phase transition $q_c$ as a function of $n$ for $\langle k \rangle=4$ $and$ $6$ with $c=0.1$. Red lines denote the solution from RA and blue lines denote the result from LA. }
\end{figure}

\subsection{Random regular (RR) NON of ER networks with clustering}

Random regular (RR) NON of ER networks with clustering indicates the internal networks are ER network with clustering and the external network of the model is a RR network with degree $m$ which means every ER network partially depends on other $m$ ER networks with clustering. The fraction of dependent nodes between every two interdependent networks is same and expressed as $q$. Assume that the initial attack is exerted on each ER network through locally removing a fraction $1-p$ of nodes from every ER network. Furthermore, each ER network has same average degree $\langle s\rangle$ and average number of triangles $\langle t \rangle$, i.e. the same size of clustering coefficient. When the cascading failures process ended, the fraction of nodes $X$ in any ER network with clustering remains which is equal to\cite{J.X.Gao2013}
\begin{equation}\label{eq11}
\begin{split}
& X=p(qYg(X)-q+1)^m,\\
& Y=p(qYg(X)-q+1)^{m-1}.\\
\end{split}
\end{equation}
where
\begin{equation}\label{eq4}
\begin{split}
& g(X)=1-e^{\langle s \rangle(1-X(1-f(X))-1)}e^{\langle t \rangle h((1-X(1-f(X)))^2-1)},
\end{split}
\end{equation}
and $f(X)=e^{\langle s \rangle(1-X(1-f(X))-1)}e^{\langle t \rangle h((1-X(1-f(X)))^2-1)}$.

The size of giant component in each network is equal to
\begin{equation}\label{eq2}
\phi_{\infty}=Xg(X).
\end{equation}

Fig.9 (a) shows the simulation results and theoretical results by comparing two sets of data $q=0.5,0.2$ and $c=0,0.15$ with $k=4$ and $m=2$. It is obvious that the simulation results agree will with theoretical results . For a given $q$, the size of the giant component of each network with larger clustering coefficient is less robust, but the gap between critical points of different $c$ is very small. For a given $c$, the phase transition changes from second order to first order with the increase of coupling strength, i.e., strong coupling strength can reduces the robustness of network. Generally speaking, the effect of coupling strength to robustness of network is more powerful than the effect of clustering coefficient. Fig.9 (b) shows the influence of coupling strength to phase transition point under localized attack and random attack. The network both become less robust with the more and more strong coupling strength. However, the critical coupling $q_c$ where the behavior of phase transition changes from first order to second order varies slightly. Fig.10 (c) and (d) shows the gap of critical threshold $p_c$ between localized attack and random attack become more and more clear with the increase of average degree $\langle k \rangle$ and clustering coefficient $c$, but their effect is same to the robustness of two different attack ways. For a fix $q$, the localized attack strength is stronger than random attack. Moreover, the robustness of network will be reduced with the increase of coupling strength and the decrease of average degree.  Fig.11 (e) shows the critical coupling $q_c$  for $\langle k \rangle=4,6$ and $c=0.1$. $q_c$ decreases  with $m$ increasing. For a fixed $m$ , the average degree $\langle k \rangle$ increase lead to the increase of $q_c$.
\begin{figure}[H] \setlength{\abovecaptionskip}{-0.05cm} \centering \scalebox{0.5}[0.5]{\includegraphics{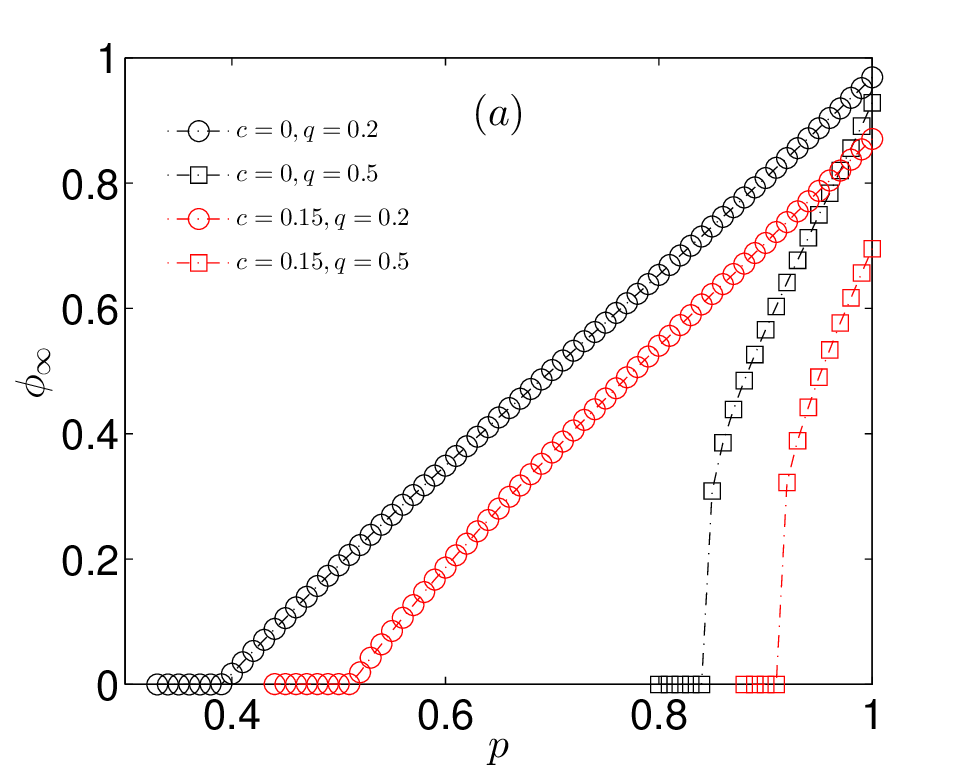}} \scalebox{0.5}[0.5]{\includegraphics{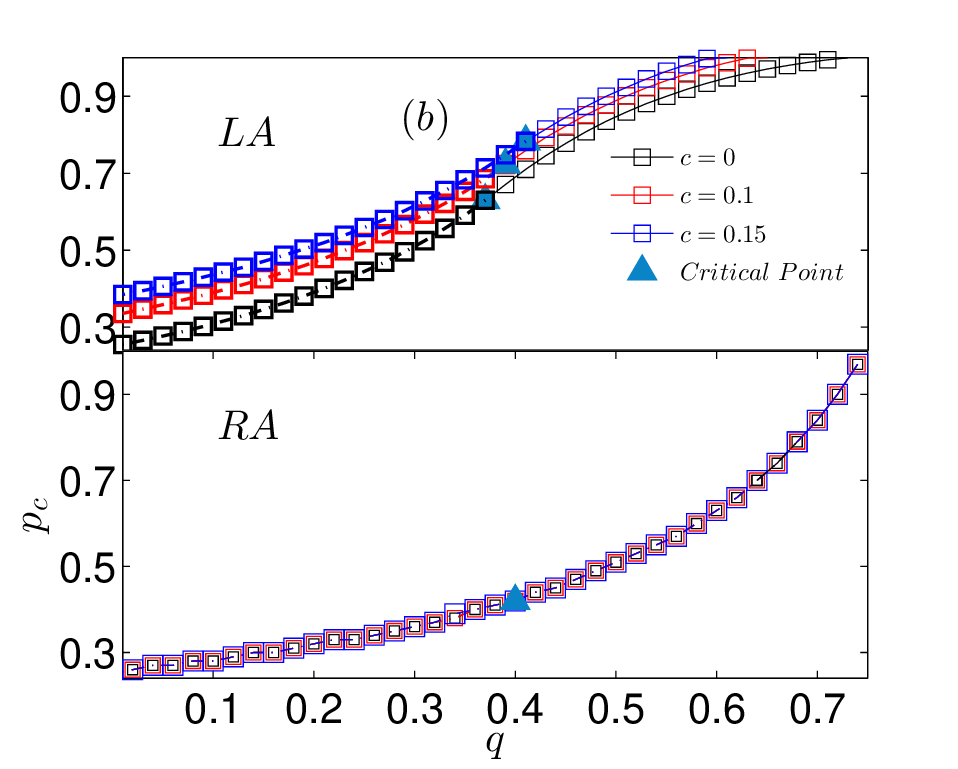}}\centering
\caption{ (a) Size of giant components $\phi_{\infty}$ as a function of $p$ for $c=0,0.1,0.15$ and $q=0.2,0.5$ with $\langle k \rangle=\langle k \rangle_A=\langle k \rangle_B=4$ and $m=2$, where dashed lines indicate the solution of theoretical predictions and symbols indicate simulations with network size $N=10000$. For strong coupling ($q=0.5$), the sizes of giant component of each network changes abruptly at critical threshold $p_c$ where shows a first order phase transition behavior. For weak coupling ($q=0.2$), the behavior is continuous, which means second order phase transition. (b) Critical threshold $p_c$ as a function of $q$ for $c=0, 0.1$ $and$ $0.15$ with $m=2$. Dashed lines denote second order phase transition (below the full blue triangle) and solid lines denote first order phase transition (above the full blue triangle). The full triangle indicates the critical point where the phase changes from second phase transition to first order phase transition.}
\end{figure}

\begin{figure}[H]
\setlength{\abovecaptionskip}{-0.05cm} \centering
\scalebox{0.5}[0.5]{\includegraphics{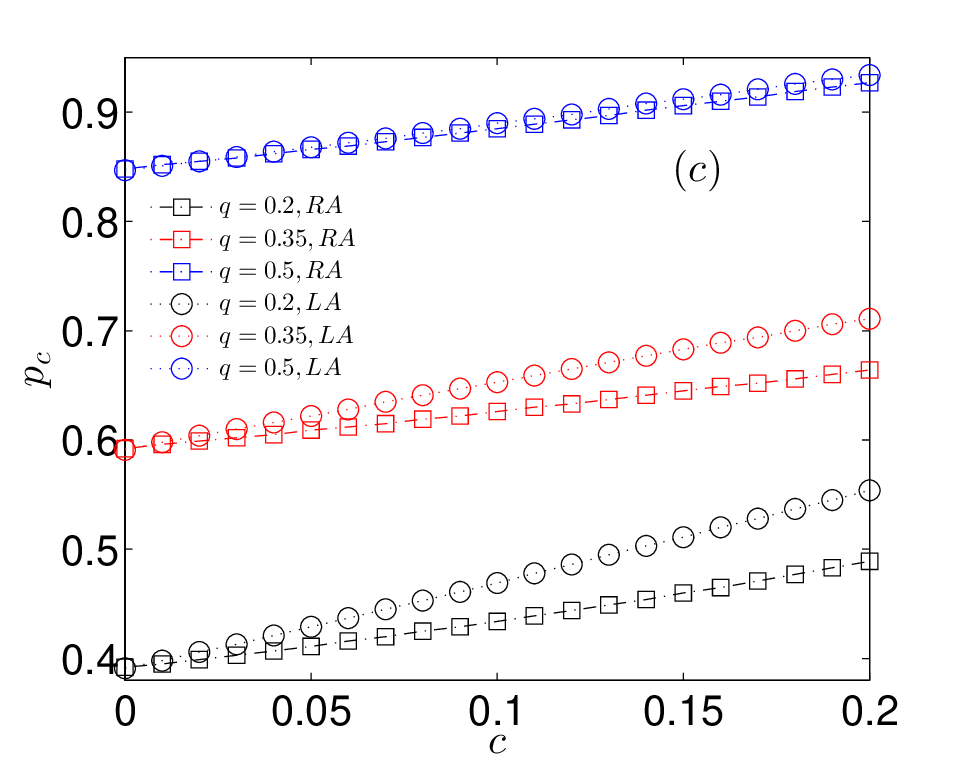}}
\scalebox{0.5}[0.5]{\includegraphics{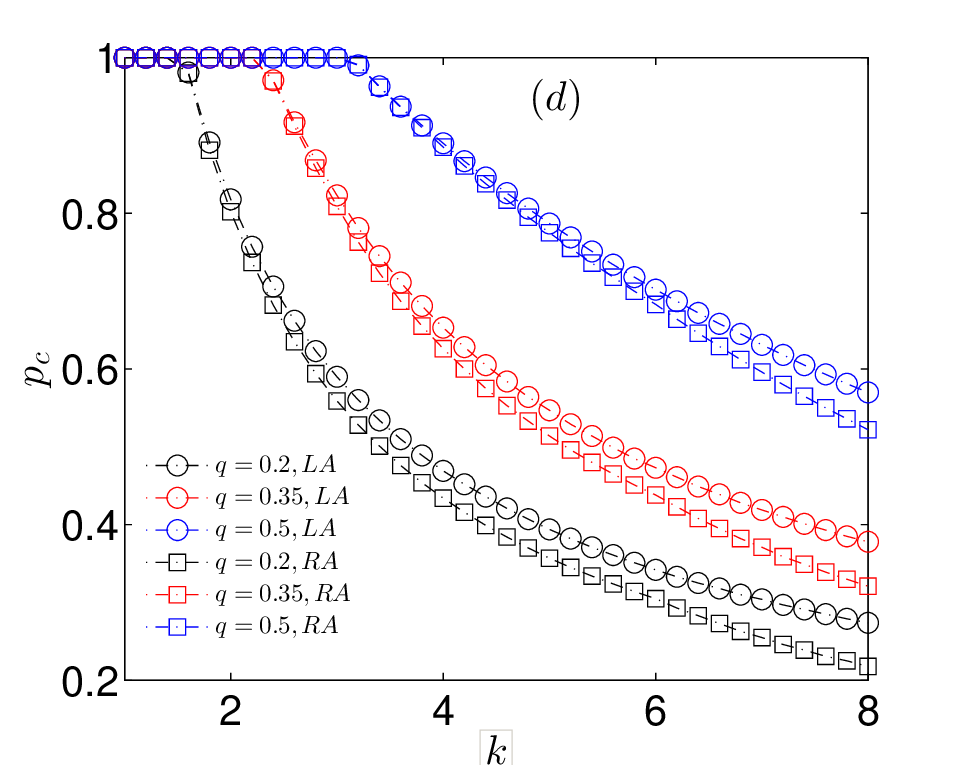}}\centering
\caption{ (c) Percolation threshold $p_c$ as a function of $c$ for $q=0.2,0.35$ $and$ $0.5$ with $\langle k \rangle=4$ and $m=2$. Circle denotes the solution from $RA$ and square denote from $LA$. (d) Critical threshold $p_c$ as a function of $k$ for $q=0.2,0.35$ $and$ $0.5$ for $c=0.1$ and $m=2$. Circle lines denote the solution from $LA$ and square denote it from $RA$. For (c) and (d) , values of strong coupling ($q=0.5$) is both bigger than weak coupling ($q=0.2$) under random attack and localized attack .}
\end{figure}

\begin{figure}[H]
\setlength{\abovecaptionskip}{-0.05cm} \centering
\scalebox{0.5}[0.5]{\includegraphics{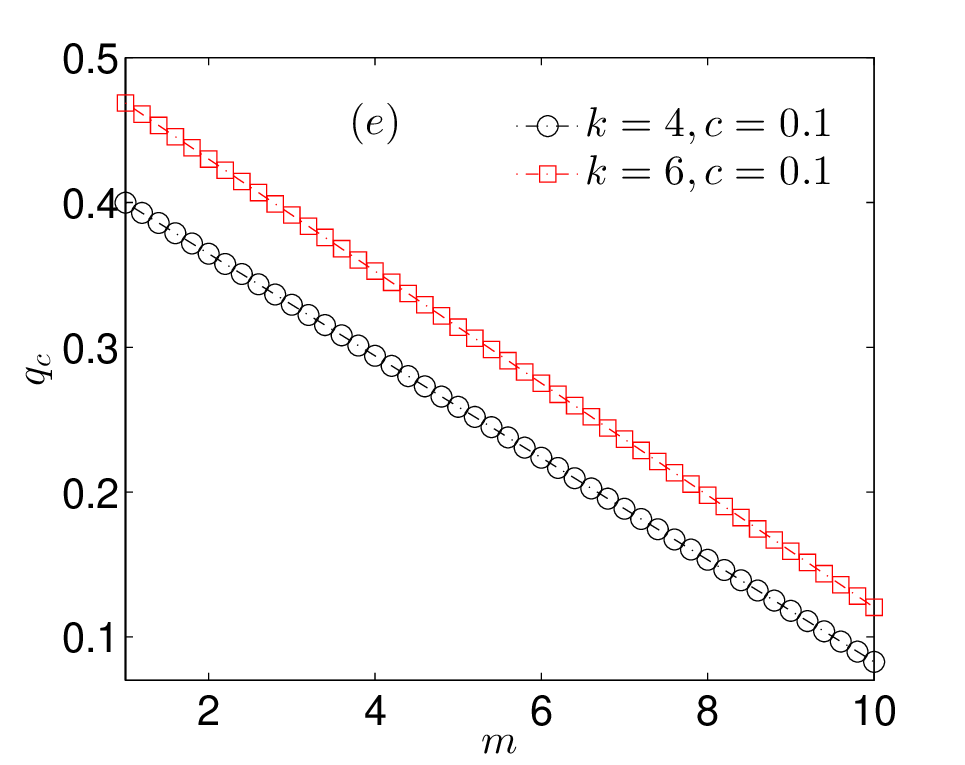}}
\centering
\caption{ (e)Critical phase transition point $q_c$ as a function of $m$ for $\langle k \rangle=4$ and $6$ with $c=0.1$. The bigger average degree has a bigger value of $q_c$. }
\end{figure}

\section{fixed degree distribution}

 Although the two interdependent clustering network that both satisfy double Poisson distribution (DPD) can obtain it's numerical results in order to analyze other features, we also get the simulation results if the two network satisfy fixed degree distribution (FDD), as shown in Fig.11.
 Fig.11 (a) and (b) show that, the robustness of network both become weak with the increase of clustering coefficient for DPD and FDD. For a fixed $c$, however, the robustness of DPD is more robust than FDD. (a) is for weak coupling strength ($q=0.3$), the behavior is continuous, i.e., second order. (b) is for strong coupling strength ($q=0.8$), the behavior is discontinuous, i.e., first order.

\begin{figure}[H]
\setlength{\abovecaptionskip}{-0.05cm} \centering
\scalebox{0.5}[0.5]{\includegraphics{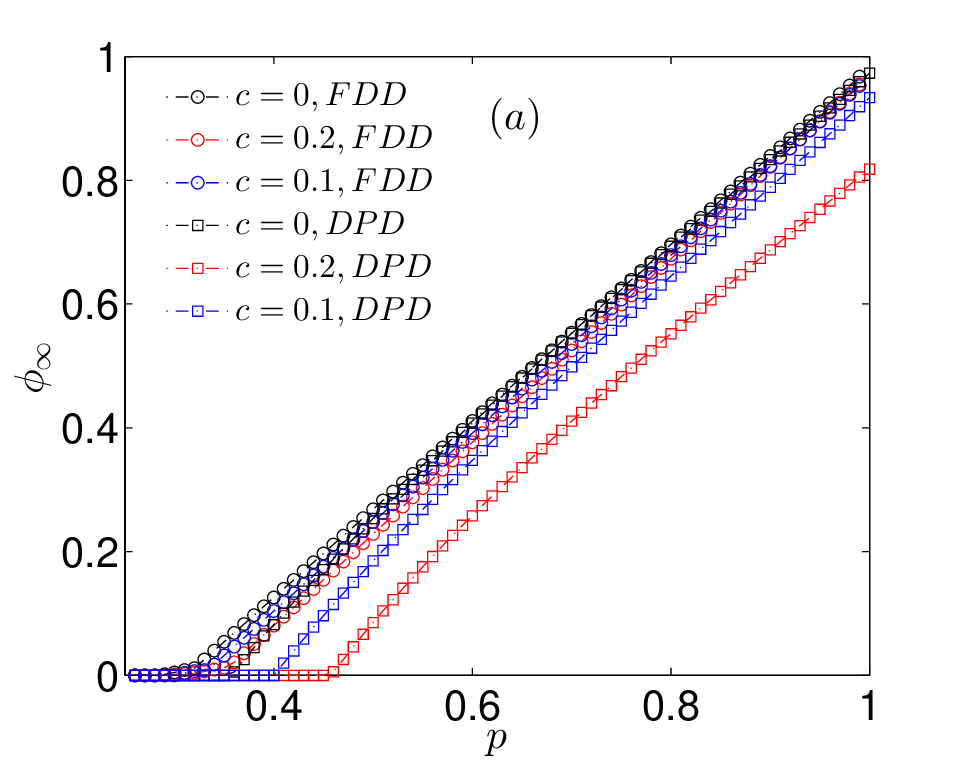}}
\scalebox{0.5}[0.5]{\includegraphics{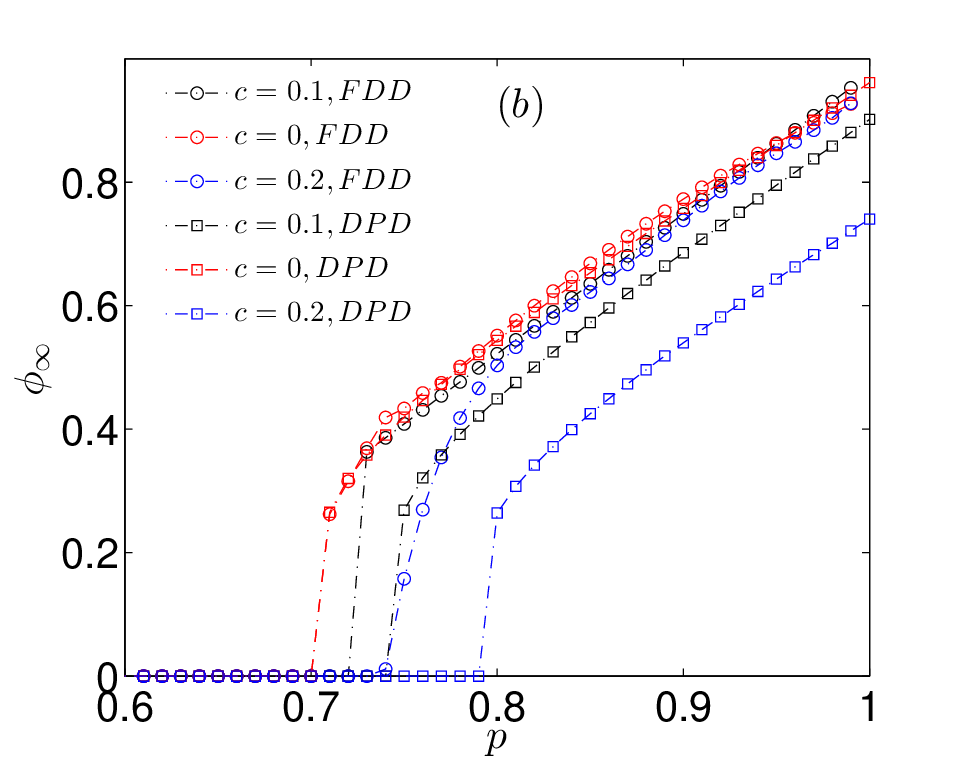}}
\centering
\caption{ Size of giant component in network A for two partially interdependent networks with clustering. Circle lines represent results for $FDD$ and circle lines represent results for $DPD$. The former results are from simulation with $N=10000$ and the later results are from numerical solution. All clustering networks degree is $\langle k \rangle=4$. (a) For weak coupling $q=0.3$, the behavior of the phase transition is second order. (b) For strong coupling $q=0.8$, the behavior of the phase transition is first order.}
\end{figure}

\section{conclusions}

By studying single clustering network, we generate the results to other network with different topological structure. For all models we have studied except the single clustered network, the other surrounding clustered networks both are partially interdependent with center clustered network. As the model shows that more stronger coupling strength lead to critical phase transition point $q_c$ increase. In the cascading process, increasing clustering coefficient $c$ result in the increase of phase transition point $p_c$, i.e., the network becomes less robust. For a certain $c$, the robustness of network become strong with the increase of average degree $k$. From weak coupling strength to strong coupling strength, the behavior of network will changes from second order to first order. In addition, we considered Star-like NON with clustering and found that higher number of clustered network $n$ causes lower value of critical phase transition point $q_c$ by comparing to various $c$ and $k$. In the last but one model, increasing average degree of RR network $m$ can causes system to be less robust. Finally, we compared the difference for different joint degree distributions (FDD and DPD) and found that the network of system obeying fixed distribution is more robust not only for strong coupling strength but also for weak coupling strength. By analyzing whole models, we can conclude that localized attack has more strong attack power than random attack.

\section{acknowledgments}
This work was supported by the National Natural Science Foundation
of China (Grant Nos. 61403171, 71403105, 71303095).

\end{document}